# AstroSat View of the Neutron Star Low-mass X-Ray Binary GX 340+0

Yash Bhargava[1], Sudip Bhattacharyya[1], Jeroen Homan[2], and Mayukh Pahari[3]
[1] Department of Astronomy and Astrophysics, Tata Institute of Fundamental Research, 1 Homi Bhabha Road, Colaba, Mumbai 400005, India
yash.bhargava_003@tifr.res.in
[2] Eureka Scientific, Inc., 2452 Delmer Street, Oakland, CA 94602, USA
[3] Department of Physics, Indian Institute of Technology Hyderabad, IITH main road, Kandi 502284, India

## Abstract

Understanding the spectral evolution along the "Z"-shaped track in the hardness–intensity diagram of Z sources, which are a class of luminous neutron star low-mass X-ray binaries, is crucial to probe accretion processes close to the neutron star. Here, we study the horizontal branch (HB) and the normal branch (NB) of the Z source GX 340+0 using AstroSat data. We find that the HB and the NB appear as two different types of X-ray intensity dips, which can appear in any sequence and with various depths. Our 0.8–25 keV spectra of dips and the hard apex can be modeled by the emissions from an accretion disk, a Comptonizing corona covering the inner disk, and the neutron star surface. We find that as the source moves onto the HB, the corona is replenished and energized by the disk and a reduced amount of disk matter reaches the neutron star surface. We also conclude that quasiperiodic oscillations during HB/NB are strongly associated with the corona and explain the evolution of strength and hard lag of this timing feature using the estimated coronal optical depth evolution.

*Unified Astronomy Thesaurus concepts:* X-ray binary stars (1811); Accretion (14); Stellar accretion disks (1579)

## 1. Introduction

A neutron star (NS) low-mass X-ray binary (LMXB), viz., an NS accreting matter from a low-mass donor star, is a natural laboratory to study accretion processes in extreme conditions. These binaries can be classified into "Z" sources and "atoll" sources based on the evolution of spectral and temporal properties, as well as luminosity (Hasinger & van der Klis 1989). Hardness–intensity diagrams (HIDs) and color–color diagrams (CCDs) of these sources provide a simple, model-independent way to probe the spectral evolution. Z sources, which emit close to the Eddington luminosity, show Z-shaped tracks in HIDs and CCDs. Such tracks can drift secularly over long duration (Hasinger & van der Klis 1989; van der Klis 2004). Z sources can be further divided into two subclasses "Cyg-like" sources (Cyg X-2, GX 340+0, and GX 5-1) or "Sco-like" sources (Sco X-1, GX 17+2, and GX 349+2), depending on the shape traced on the HID (Kuulkers et al. 1994; Homan et al. 2010). RXTE observations of the transient sources XTE J1701−462 and IGR J17480−2446, evolved through both Z and atoll phases, have shown that the atoll and Z source (sub)classes are probably strongly related to the mass accretion rate (Homan et al. 2007; Lin et al. 2009; Homan et al. 2010; Chakraborty et al. 2011).

Z tracks, usually studied in ∼2–20 keV band, can be subdivided into three branches; horizontal branch (HB), normal branch (NB), and flaring branch (FB; Hasinger & van der Klis 1989; van der Klis 2004). Since the source typically moves along a single track, this motion can be parameterized by a single parameter $S_z$, which increases from the HB, via the NB, to the FB (with the HB/NB vertex (hard apex) set to $S_z = 1$ and the NB/FB vertex (soft apex) set to $S_z = 2$). However, the nature of the changes along the Z track and what drives them is still not understood. Hasinger et al. (1990) suggested that the evolution is driven by changes in the mass accretion rate, increasing from the HB to the FB, while Church et al. (2006) argued that the mass accretion rate increased from the NB to HB, with the FB being the result of changes in the thermonuclear burning rate. Homan et al. (2002) suggested that the mass accretion rate may not change much at all along the Z track, and Lin et al. (2009) proposed that motion along the Z track could arise due to instabilities in the accretion disk.

Z tracks suggest a distinctive and repetitive spectral evolution of Z sources. Therefore, an explanation of such tracks should be a key to understand the accretion processes of this class of bright NS LMXBs. The spectral evolution of Z sources has often been studied in terms of transitions from the FB to the NB and HB (and vice versa) with the main contribution to the 2−20 keV spectrum arising from a thermal and a nonthermal component (e.g., Mitsuda et al. 1989; Church et al. 2006, 2012). According to Mitsuda et al. (1989), the soft thermal component arises due to a multicolor (disk) blackbody, while Church et al. (2006), Jackson et al. (2009), and Bałucińska-Church et al. (2010) preferred it to be a blackbody-like component, possibly from the boundary layer on the NS surface. The nonthermal component was assumed to arise from the Comptonized emission from a hot electron cloud, often referred to as a corona. But at softer energies (0.5−2 keV), another component was necessary to describe the spectrum (Lavagetto et al. 2004; Seifina et al. 2013). The spectrum of Z sources also has been interpreted as a combination of a thermal accretion disk, blackbody component, and the Comptonized emission in which the both disk and blackbody are relatively hotter (∼1.5 and ∼2.4 keV respectively) and the Comptonized emission is weaker with a break at ∼20 keV (Homan et al. 2007; Lin et al. 2007, 2009, 2012; Homan et al. 2010).

Along the Z track, HB shows kHz quasiperiodic oscillations (QPOs), HB oscillations (HBOs) in ∼25–50 Hz, and a low-frequency broadband noise, and NB shows NB oscillations (NBOs) in ∼5–7 Hz (van Paradijs et al. 1988; Penninx et al. 1991; Jonker et al. 1998, 2000; Sriram et al. 2011). While various models have been proposed to explain these oscillations, their origins are still unclear (e.g., van der Klis 2004).

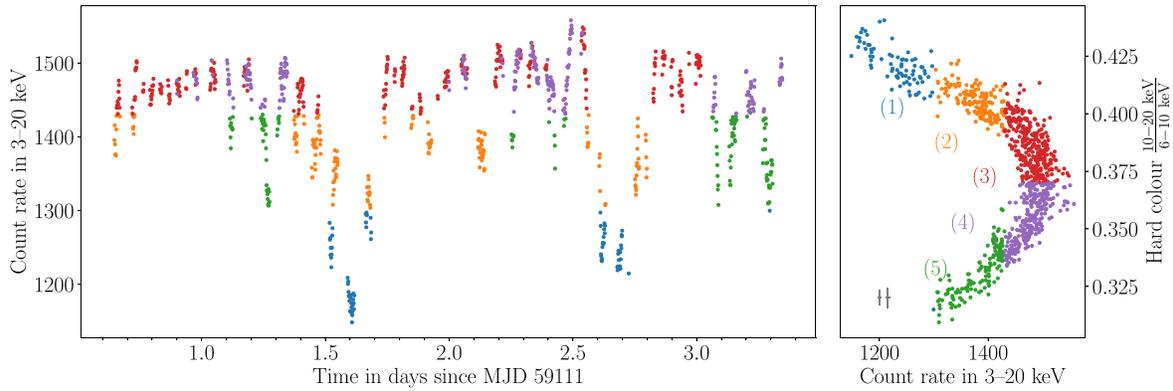

**Figure 1.** Left: AstroSat/LAXPC (LXP 20) light curve of GX 340+0 in 3–20 keV with 100 s bins. Right: the LXP 20 hardness–intensity diagram (HID) of the source with hard color computed using the background-subtracted count rates in 10–20 keV divided by that in 6–10 keV and intensity represented by the total count rate in 3–20 keV. Representative error bars (minimum and maximum) are shown in gray at the bottom left corner. A part of the Z track (HB and NB) is clearly seen and has been divided into five zones with roughly similar extent on the Z track. Different colors have been used for points corresponding to different zones. The light curve of the left panel is coded with the same colors, which show how the source moved from one zone to another on the Z track.

GX 340+0 is one of the brightest and well-studied Z sources in the Galactic plane (Mitsuda et al. 1989; Jonker et al. 1998, 2000; Gilfanov et al. 2003; Lavagetto et al. 2004; Iaria et al. 2006; Bałucińska-Church et al. 2010; Seifina et al. 2013). It shows a clear and repetitive Z track with strong detections of HBOs and NBOs (Jonker et al. 1998; Gilfanov et al. 2003; Bałucińska-Church et al. 2010). The source can trace out a full Z track, i.e., covering all the branches, on timescales as short as a few days (Jonker et al. 2000). The source is highly absorbed with the neutral hydrogen column density reported in the range $(6-12) \times 10^{22}$ cm$^{-2}$ (Church et al. 2006; Iaria et al. 2006). Mitsuda et al. (1989) described the soft thermal component as an accretion disk, while Church et al. (2006) described it as a single-temperature blackbody. Both these prescriptions also required a nonthermal component that was modeled as a Comptonized emission. However, Seifina et al. (2013) described the broadband spectrum from BeppoSAX as a combination of a low-temperature blackbody (from the accretion disk) and two Comptonization components (the softer one from the transition layer between the disk and the NS and the harder one from the NS surface). Church et al. (2006) explained the HB in terms of the reduction of blackbody area and Comptonized emission and an increase of the blackbody temperature. However, Seifina et al. (2013) attributed the HB evolution to the change of the electron temperature of a Comptonizing transition layer.

Thus, even decades after the first studies of Z tracks, the cause of these distinctive tracks is not reliably known. In this article, we attempt to explain the HB and the NB of GX 340+0 by analyzing the AstroSat light curve and broadband spectra.

## 2. Observation and Data Reduction

AstroSat (Singh et al. 2014) observed the GX 340+0 from 2020 September 19 to 2020 September 22 (observation ID A09_134T01_9000003896), corresponding to a total LAXPC exposure of 92.8 ks and a total SXT exposure of 29.6 ks. The observation covered HB and NB phases of the Z track as seen in the right panel of Figure 1. The AstroSat observation was conducted simultaneously with Large Area X-ray Proportional Counter (LAXPC) in event analysis mode (Yadav et al. 2016, 2017) and Soft X-ray Telescope (SXT) in photon counting mode (Singh et al. 2016, 2017). LAXPC has three independent proportional counters, but one of them suffered a gas leak early in the mission (LXP 30), and another has shown abnormal gain variations since early 2018 (LXP 10; Antia et al. 2021). Thus we use only LXP 20 for the analysis. We reduce the LAXPC data using the pipeline LAXPCSOFTWARE22AUG15 (Antia et al. 2021; Misra et al. 2021).[4] The software includes tools to reduce the level 1 data to level 2 data, calibration files, responses, etc. We obtain the level 2 data and standard good time intervals (which exclude the duration of South Atlantic Anomaly passage, etc.) using the tools from the above-mentioned pipeline. We also extract the energy-dependent light curves, spectra, background spectra, and background light curves using the tools available in LAXPCSOFTWARE22AUG15. We consider the data from layer 1 of the instrument due to its minimal background contribution.

The processed SXT level 2 files were downloaded from Astrobrowse.[5] We extract the standard products (energy-dependent light curves and spectra) using HEASARC (version 6.30.1) tool xselect. The source count rate is too low to have significant pileup effects, and thus a circular region of 15′ is used to extract the data products. We use the standard background and response files provided by the SXT payload operation center (POC).[6] We modify the standard ancillary response file to correct for the smaller source region using the tool provided by the POC.

## 3. Spectral and Timing Analysis

### 3.1. Demarcation of the HID Zones

We investigate the evolution of the source by extracting energy-dependent light curves (with 100 s bins) and mapping the count-rate evolution to positions on the HID (see Figure 1 for description). We divide the HID into five different zones, ensuring roughly similar length is covered by each zone on the track (see Figure 1). Zone 1 and 2 correspond to the HB, Zone 3 and 4 correspond to the hard apex, and Zone 5 corresponds to

---

[4] http://astrosat-ssc.iucaa.in/uploads/laxpc/LAXPCsoftware22Aug15.zip
[5] https://astrobrowse.issdc.gov.in/astro_archive/archive/Home.jsp
[6] https://www.tifr.res.in/~astrosat_sxt/index.html

Table 1
Spectral and Timing Parameter Values for All HID Zones of GX 340+0

| Component | Parameter | Units | Zone ($S_z$ range) | | | | |
|---|---|---|---|---|---|---|---|
| | | | 1 (0.5–0.66) | 2 (0.66–0.83) | 3 (0.83–1) | 4 (1–1.25) | 5 (1.25–1.5) |
| tbabs | $n_H$ | cm$^{-2}$ | $10^{23a}$ | | | | |
| diskbb | $T_{in}$ | keV | $0.24 \pm 0.02$ | $0.28 \pm 0.02$ | $0.28 \pm 0.01$ | $0.30^{+0.01}_{-0.03}$ | $0.27 \pm 0.02$ |
| | Norm[b] | $10^5$ | $9.1^{+12.4}_{-3.4}$ | $2.4^{+1.5}_{-0.8}$ | $3.3 \pm 0.9$ | $2.1^{+2.0}_{-0.4}$ | $3.4^{+2.5}_{-1.3}$ |
| | Flux (3–20 keV)[c] | | $(7 \pm 1) \times 10^{-4}$ | $(25 \pm 3) \times 10^{-4}$ | $(28 \pm 3) \times 10^{-4}$ | $(40 \pm 4) \times 10^{-4}$ | $(26 \pm 5) \times 10^{-4}$ |
| bbodyrad | $T_{BB}$ | keV | $1.14^{+0.04}_{-0.02}$ | $1.15^{+0.01}_{-0.02}$ | $1.13 \pm 0.01$ | $1.14^{+0.02}_{-0.01}$ | $1.14 \pm 0.01$ |
| | Norm[d] | | $356^{+40}_{-64}$ | $449 \pm 50$ | $622 \pm 49$ | $638^{+24}_{-107}$ | $592 \pm 47$ |
| | Flux (0.01–100 keV)[c] | | $0.65 \pm 0.01$ | $0.85 \pm 0.01$ | $1.10 \pm 0.01$ | $1.18 \pm 0.01$ | $1.07 \pm 0.01$ |
| | Flux (3–20 keV)[c] | | $0.45 \pm 0.01$ | $0.59 \pm 0.01$ | $0.75 \pm 0.01$ | $0.81 \pm 0.01$ | $0.73 \pm 0.01$ |
| Nthcomp | $\Gamma$ | | $1.76^{+0.11}_{-0.06}$ | $1.66 \pm 0.08$ | $1.51 \pm 0.10$ | $1.45^{+0.23}_{-0.07}$ | $1.66 \pm 0.12$ |
| | $kT_e$ | keV | $3.21 \pm 0.08$ | $3.14 \pm 0.07$ | $2.98 \pm 0.07$ | $2.91^{+0.22}_{-0.05}$ | $3.05 \pm 0.10$ |
| | Flux (0.01–100 keV)[c] | | $1.26 \pm 0.04$ | $1.08 \pm 0.04$ | $0.86 \pm 0.03$ | $0.69 \pm 0.04$ | $0.79 \pm 0.05$ |
| | Flux (3–20 keV)[c] | | $0.65 \pm 0.01$ | $0.63 \pm 0.02$ | $0.57 \pm 0.01$ | $0.48 \pm 0.02$ | $0.45 \pm 0.02$ |
| | Total Flux (3–20 keV)[c] | | $1.10 \pm 0.06$ | $1.23 \pm 0.05$ | $1.33 \pm 0.04$ | $1.30 \pm 0.03$ | $1.19 \pm 0.03$ |
| | $\chi^2$ | | 93.5 | 91.3 | 128.2 | 106.7 | 83.6 |
| | DoF | | 104 | 107 | 114 | 109 | 102 |
| | Timing parameters[e] | | | | | | |
| | QPO frequency | Hz | $29.3 \pm 0.1$ | $36.8 \pm 0.1$ | $43.5^{+0.5}_{-0.2}$ | $48.5^{+0.3}_{-0.5}$ | $5.2 \pm 0.1$ |
| | QPO FWHM | Hz | $6.7 \pm 0.2$ | $7.3 \pm 0.4$ | $15^{+0.5}_{-1.5}$ | $9.3^{+2.7}_{-0.8}$ | $2.3^{+0.4}_{-0.7}$ |
| | Fractional rms | % | $7.2 \pm 0.1$ | $5.1 \pm 0.1$ | $4.18^{+0.05}_{-0.18}$ | $2.2 \pm 0.1$ | $2.4^{+0.1}_{-0.4}$ |
| | Phase lag | radian | $0.11 \pm 0.05$ | $0.79 \pm 0.18$ | $3.05 \pm 0.02$ | ... | ... |
| | Time lag | ms | $0.58 \pm 0.25$ | $3.43 \pm 0.77$ | $11.15 \pm 0.14$ | ... | ... |

**Notes.** AstroSat SXT+LAXPC spectra are fitted with the XSPEC model tbabs*(diskbb+bbodyrad+Nthcomp), and the Nthcomp seed photon temperature is tied to the diskbb temperature. 1σ errors are shown.
[a] The parameter is frozen to the value.
[b] The normalization is defined as $(R_{in}/D_{10})^2 \cos(\theta)$, where $D_{10}$ is source distance in units of 10 kpc, $R_{in}$ is apparent inner disk radius in kilometers, and $\theta$ is the disk inclination angle.
[c] Unabsorbed flux in $10^{-8}$ erg cm$^{-2}$ s$^{-1}$ as estimated using cflux on the corresponding component.
[d] The normalization is defined as Norm = $(R/D_{10})^2$, where $R$ implies the size of the blackbody region in kilometers.
[e] The timing parameters are derived by fitting the PDS (Figure 4) with phenomenological models and characterizing the oscillation with a Lorentzian function. The lags are computed between 3–7 keV and 7–20 keV with positive lags indicating that the hard band is lagging the soft band.

the NB. We map the zones to the corresponding points in the light curve using different colors to probe the evolution of the source on the Z track. To extract the HID zone–resolved products, we identify the observed time intervals of each zone. These intervals are used to extract the simultaneous energy spectra from SXT and LAXPC (see Section 3.2) and power density spectra from LAXPC observations (see Section 3.3). To determine the typical $S_z$ range covered by our observation, we use the timing properties of the source as seen in the power density spectra (PDS; see Section 3.3) and compare it with the source properties as seen by Jonker et al. (2000). The HBO frequency of zone 1 would indicate that the $S_z \approx 0.6$ for zone 1 and transition from HBO to NBO (which happens in Zone 4–5) would place their $S_z$ at $\approx 1.4$. Hence we assume that the source is observed at $S_z$ in the range 0.5−1.5 and divide the interval in Table 1 accordingly.

### 3.2. Spectral Analysis of the HID Zones

We model[7] the simultaneous SXT and LAXPC spectra jointly (with total exposure for all HID zones with LAXPC at 92.8 ks and SXT at 29.6 ks), using the 0.8–7 keV range for SXT (Chakraborty et al. 2020) and the 4–25 keV range for LAXPC (the background dominates above ∼25 keV; Bhargava et al. 2019). The spectra of the source are shown in Figure 2. The spectra from both instruments are rebinned to match the respective spectral resolutions. We include 1.5% systematic error in SXT spectra (Sridhar et al. 2019) and 3% systematic error in LAXPC spectra (Bhargava et al. 2019). We also note that the spectrum of different HID zones have a distinct spectral shape, which is evident in the ratio of all the zone spectra to the spectra from Zone 3 (see Figure 2 bottom left panel).

To characterize the spectrum, we first note that the source is highly absorbed (Church et al. 2006; Iaria et al. 2006). We initially model the spectrum with an absorbed cutoff power law (cutoffpl). We include the neutral hydrogen absorption using the XSPEC model tbabs with abundances as suggested by Wilms et al. (2000) and the cross sections from Verner et al. (1996). We find that the spectrum of the source is consistent with a column density of $10^{23}$ cm$^{-2}$ (Church et al. 2006; Iaria et al. 2006). Similar to Mitsuda et al. (1989), Lavagetto et al. (2004), Church et al. (2006), Lin et al. (2012), and Seifina et al. (2013), we find that the spectrum requires at least an additional thermal component. For example, for zone 3, $\chi^2 \approx 979$ for

---

[7] Using the standard XSPEC software (version 12.12.1): https://heasarc.gsfc.nasa.gov/xanadu/xspec/.

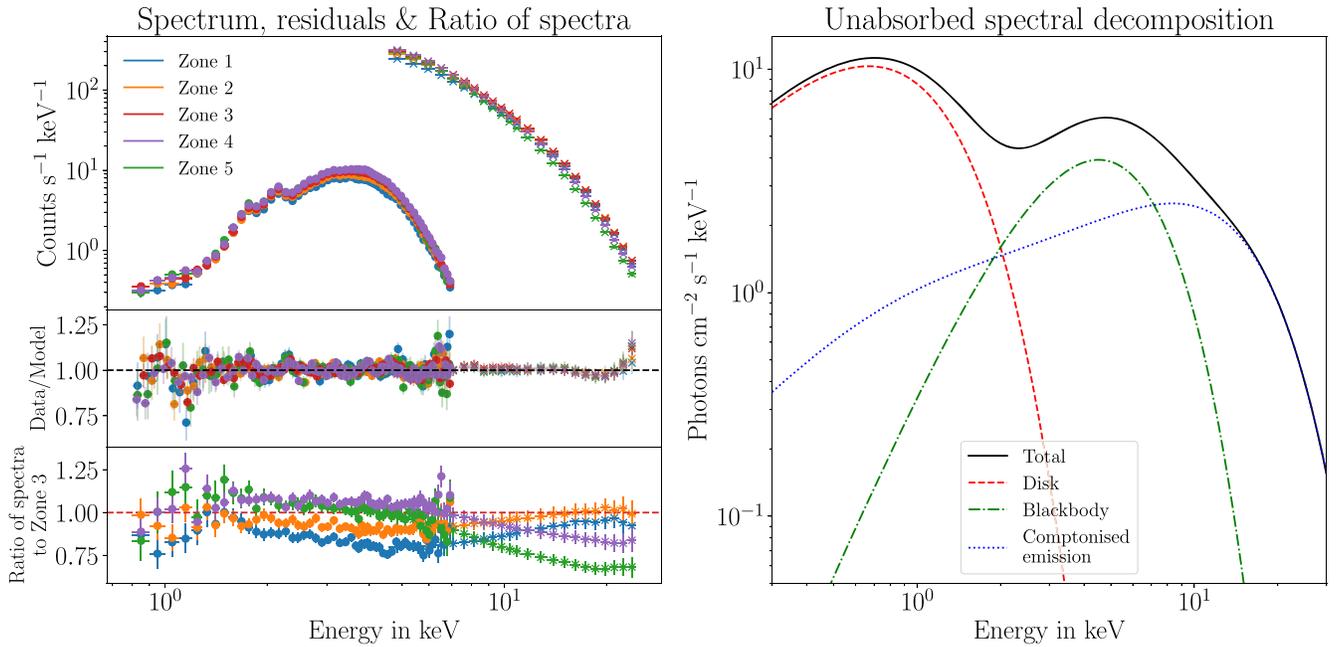

**Figure 2.** Top left: HID zone–resolved spectra for GX 340+0 as extracted from AstroSat SXT (circles) and LAXPC (crosses). The spectra for different HID zones are colored according to the scheme shown in Figure 1. Middle left: the residuals from the best-fit model (Table 1) are shown for each HID zone. Bottom left: ratio of the zone-wise spectra to Zone 3. Right: the XSPEC spectral model tbabs*(diskbb+bbodyrad+Nthcomp) for best-fit parameter values for zone 2 with various components (see Table 1 for best-fit parameter values and Section 3.2 for more details). For the decomposition plot, the absorption was set to 0.0 to highlight the relative contribution of each component.

118 degrees of freedom (DoF) for an absorbed power law (tbabs*cutoffpl in XSPEC) improves to $\chi^2 \approx 294$ for 116 DoF for an absorbed cutoff power law plus a blackbody (tbabs*(cutoffpl+bbodyrad) in XSPEC). However, the fitting is still not adequate as there is a significant residual at lower energies. We model this additional component with a multicolor-disk blackbody, diskbb, which gives an acceptable $\chi^2 \approx 128$ for 114 DoF. Thus, this spectral model comprises two thermal components and a phenomenological cutoff power law component. Here, diskbb and bbodyrad are the softer (∼0.3 keV) and the harder (∼1 keV) thermal components, respectively. Note that $\chi^2$ increases from 128 to 148 (DoF =114) if we interchange the temperatures of diskbb and bbodyrad and determine the best fit.

To have a physical picture of the nonthermal component, we replace the cutoffpl with the thermal Comptonization model Nthcomp (Zdziarski et al. 1996; Życki et al. 1999), with the seed photons provided by the soft disk. Hence, we tie the diskbb temperature with the seed photon temperature of Nthcomp. Thus, we use the XSPEC model constant*tbabs*(diskbb+bbodyrad+Nthcomp) (Lin et al. 2009, 2012), which works well for all HID zones (see left panel of Figure 2 and Table 1). We fix the absorption column density $n_H$ to $10^{23}$ cm$^{-2}$ (Mitsuda et al. 1989; Church et al. 2006; D'Aì et al. 2009; Seifina et al. 2013; Miller et al. 2016) for all the zones and find acceptable fits. Note that a free $n_H$ does not improve the fits significantly. Moreover, the trends observed in parameters from the fits are independent of the value at which $n_H$ is fixed. The evolution of the key parameters is shown in Figure 3, while the best-fit spectral parameters are noted in Table 1. The optical depth of the Comptonizing corona can be computed from Nthcomp index Γ and electron temperature $kT_e$ (Bhargava et al. 2019).

### 3.3. Timing Analysis

We extract the PDS in 0.03–500 Hz for all HID zones mentioned in Section 3.1. For the extraction of the PDS we use a custom software GHATS.[8] The PDS are shown in Figure 4. This figure shows HBOs for zones 1–4 in ∼30–50 Hz and NBOs for zone 5 in ∼5 Hz. Note that the PDS of zone 4 also indicates a broad feature at NBO frequencies. The PDS is converted to XSPEC readable format and is modeled with multiple Lorentzian functions (to characterize broad-noise features and QPOs) and a constant (for the white noise). Properties of these timing components, such as QPO peak frequency, FWHM, and strength (fractional rms amplitude), are estimated from modeling. We also extract the PDS in various energy bands (in 3–4.5, 4.5–6, 6–10, and 10–20 keV) to trace the dependence of the QPO fractional rms amplitude on energy. We show the energy-dependence of QPO fractional rms amplitude in Figure 5. In addition, we compute the phase and time lags between 3–7 keV and 7–20 keV bands for strong QPOs (i.e., HBOs in zones 1–3) by averaging over the QPO FWHM (the lags are reported in Table 1).

The timing features show a strong dependence on the position on the HID. To gain an insight about how these features evolve with time, we depict the frequency–time image of GX 340+0 in Figure 6. We also overplot the 3–20 keV light curve to highlight the correlated evolution of count rates and the peak frequency of timing features. As the source goes onto the HB, the HBO frequency decreases, and the feature becomes stronger. The opposite happens when the source comes out of the HB. The low-frequency broadband noise becomes strong when the HBO frequency goes below ∼30 Hz. This is consistently seen for two deep HBs, as well as for the shallow HB.

---

[8] Developed by Tomaso Belloni.

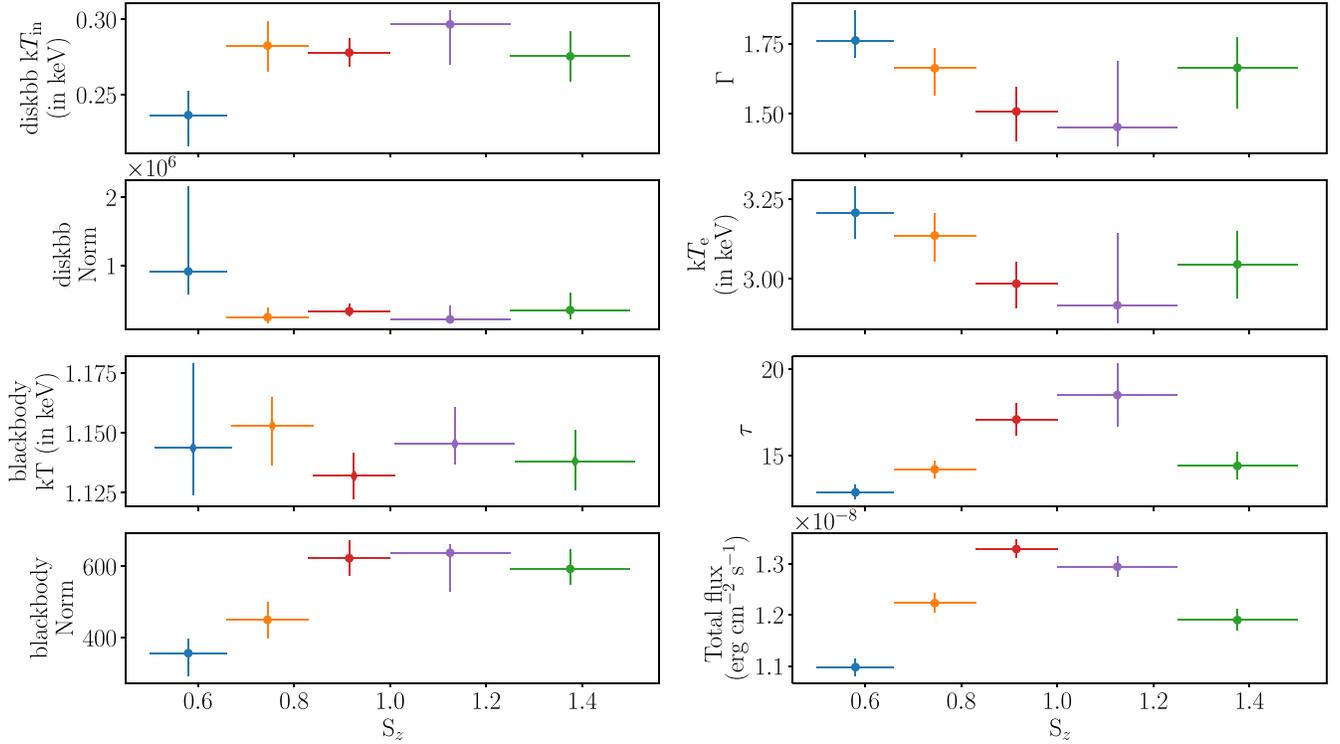

**Figure 3.** HID zone dependence of key spectral parameters from the XSPEC model tbabs*(diskbb+bbodyrad+Nthcomp) for the SXT+LAXPC joint modeling of GX 340+0 spectra. The best-fit parameter values corresponding to the disk and blackbody components are shown in the left panels and those of the Comptonizing corona are depicted in the right panels. The total flux depicted in bottom right panel corresponds to the total unabsorbed flux in 3–20 keV. The variation in the blackbody normalization and the complex evolution of the Comptonizing corona lead to the movement of the source along the Z track.

## 4. Results and Discussion

### 4.1. Sampling of the Z Track

Using observations of GX 340+0 with AstroSat, we investigate source spectral and timing properties to understand its evolution along the Z track. Our observation traces parts of the HB and the NB several times without going into the FB. The observation spanned roughly 3.5 days, which is similar to the timescale on which the source has been found to trace out a complete Z track in the past (Jonker et al. 2000). This shows that the timescale on which full Z tracks are traced out can vary in GX 340+0. Similar behavior was observed in the Z source GX 17+2 by Lin et al. (2012). In their observation, the source first lingered on the HB and NB for about 6 days and before tracing out a full Z track within ∼1.5 days.

### 4.2. HB and NB as the Source Excursions into Dips

In our observation, we find that GX 340+0 spends most of its time in HID zones 3 and 4 (see Figure 1) and occasionally makes excursions into HB and NB, which appear as dips in the light curves. The appearance of HB and NB excursions as dips in the light curves is a property that is common for Z sources in their Cyg-like Z phase, like GX 340+0. As can be seen for the Z sources analyzed in Fridriksson et al. (2015), as the sources evolve from Sco-like Z source behavior to Cyg-like Z source behavior, the count-rate variations along the HB and NB become increasingly stronger. The dips in GX 340+0 are of two distinctly different types: with high hard-color values (>0.375) implying the HB and with low hard-color values implying the NB. For our LAXPC observation, the source is observed around the hard apex, the HB, and the NB for 60.5, 21.2, and 11.1 ks, respectively (excluding data gaps). The depth of the dips, corresponding to the length of the HB or the NB, can also vary apparently randomly. For example, Figure 1 shows shallow HB dips when the source goes up to zone 2 and comes back to the hard apex. In our observation, there are two deep and longer (∼25 ks, including the data gaps) HB dips, but all NB dips are relatively shallow and short (∼10 ks, including the data gaps). Note that when the source goes into a dip it gradually comes out to the hard apex in the same way, without jumping to another dip or branch (also seen in other Z sources; Homan et al. 2007; Sriram et al. 2011; Lin et al. 2012, etc.). This indicates well-defined changes in the accretion processes during a dip or a Z-track branch. Therefore, we analyze spectral properties during dips to probe the accretion processes.

### 4.3. Understanding the Shape of the Z Track

A long-standing goal regarding Z sources is to understand the shape of their tracks in the HID, particularly the vertices. Thus, before describing detailed spectral properties, we attempt to reproduce the Z track of GX 340+0 using our spectral components. This track appears in 3–20 keV (Figure 1), and only bbodyrad and Nthcomp significantly contribute in this energy range (Table 1). Thus, relative flux values of bbodyrad and Nthcomp (see right panel of Figure 2) are expected to give rise to the Z track of Figure 1. To test this, we compute the Nthcomp flux to bbodyrad flux ratio in the same energy range (6–20 keV) in which the hard color is computed. We also estimate the total flux in the same energy range (3–20 keV) as used to compute the intensity of HIDs (Figure 1). In Figure 7, we plot this flux ratio versus total flux. This plot qualitatively reproduces (including the bend) the Z track of GX 340+0 (see Figure 1), as shown in the left panel of Figure 7. This implies our spectral decomposition can describe the motion of the source on the HID and explains how the HB and the NB originate from the evolution of and the interplay between blackbody and Comptonizing components.

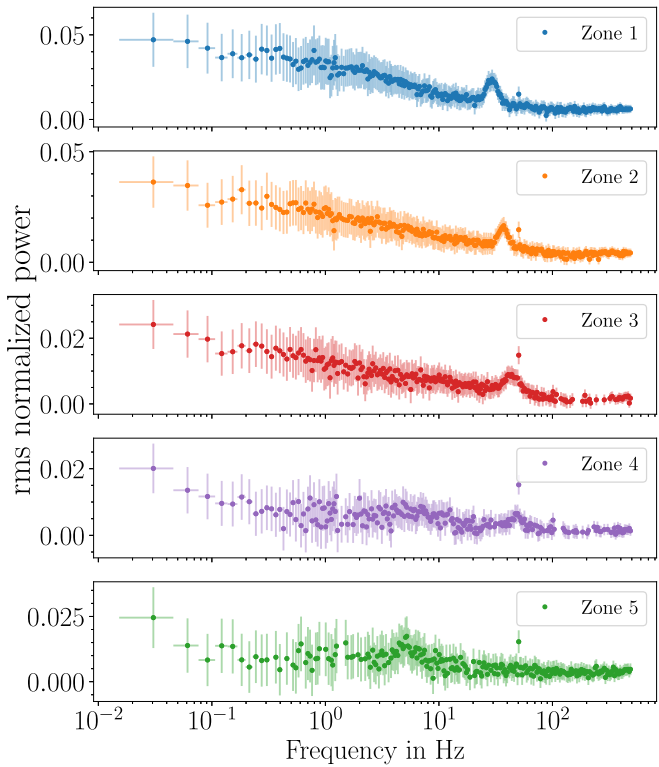

**Figure 4.** The power density spectrum for each HID zone with the same color scheme used in Figure 1. The excess power in the single bin at the 50 Hz is instrumental. HBOs and NBO are seen in zones 1–4 and zone 5, respectively (see Table 1).

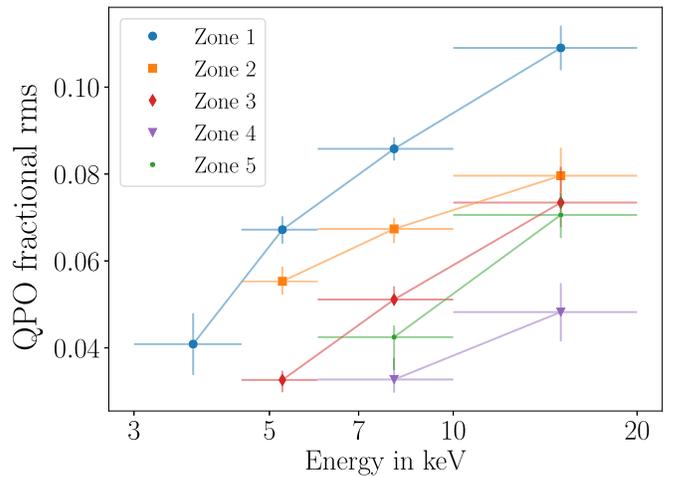

**Figure 5.** The QPO fractional rms amplitude vs. energy plot for all HID zones (HBOs for zones 1–4 and NBO for zone 5). QPOs are detected for all four energy bands in 3–20 keV only for zone 1.

### 4.4. Spectral Components and Evolution of Spectral Parameters

The three components, accretion disk, NS surface, and Comptonizing corona, used in our X-ray spectral analysis, are naturally expected for NS LMXBs (Lin et al. 2007; Mukherjee & Bhattacharyya 2011; Lin et al. 2012). All three components are significant (see Section 3.2), which is a progress relative to the previous two-component models (see Section 1).

Our three-component spectral model adequately fits spectra of the hard apex and HB/NB dips of GX 340+0 (see Section 3.2). This enables us to uniformly study the observed HB and NB using the evolution of spectral parameters. This evolution can be seen in Figure 3 and Table 1. We find that the disk inner edge is far from the NS in the HID zone 1 ($\sim$1100 km)[9] and may be somewhat nearer for other zones (500 $-$700 km). As expected, the trend of the diskbb temperature ($kT_{in}$) is opposite to that of diskbb normalization (Frank et al. 2002). The disk temperature is consistent with the estimates from Seifina et al. (2013) but slightly lower than typical Z-source disk temperature (e.g., GX 17+2; Lin et al. 2012). The blackbody component could naturally be explained in terms of a spreading layer of accreted matter on the NS surface. This implies that the disk material reaches the NS. Moreover, the Keplerian disk is believed to extend up to near the NS surface for a high-luminosity NS LMXB like GX 340+0. But a large-disk inner radius (see Table 1) means that the inner part of the disk is possibly hidden. This indicates that the corona covers the inner disk and hence gets seed photons from the disk. We find that the blackbody temperature remains almost constant throughout the HID zones. But its normalization, which can be interpreted as proportional to the spreading layer area on the NS surface, increases from zone 1 to zone 3 (i.e., from the HB to the hard apex) and then remains almost the same in the hard apex and the NB. The blackbody bolometric (0.01–100 keV) flux also increases from zone 1 to zone 3. But all of Nthcomp index ($\Gamma$), electron temperature ($kT_e$), and bolometric flux decrease, while the optical depth ($\tau$) increases, from zone 1 to zone 3. The right panel of Figure 7 shows that the trend of the bolometric flux of the blackbody is exactly opposite to that of the Comptonizing corona. In addition, Table 1 shows that the decrease in the blackbody flux causes the HB dip in the 3–20 keV range.

### 4.5. Spectro-timing Properties

HID zones 1–4 show a low-frequency broad noise and HBOs, while zone 5, and perhaps also zone 4, show NBOs (Figure 4). Note that simultaneous observations of HBOs and NBOs (such as indicated for our zone 4) were previously reported for various Z sources (Homan et al. 2002; Jonker et al. 2000, 2002; Sriram et al. 2011). Our timing analysis shows HBOs ($\sim$30–50 Hz) and NBOs ($\sim$5 Hz) are observed in the expected frequency ranges (e.g., van Paradijs et al. 1988; Penninx et al. 1991; Jonker et al. 2000, see also Section 1). Table 1 shows that the HBO frequency decreases but the strength increases, when the source goes into the HB dip (i.e., zone 4 to zone 1; see also Figures 4 and 6). This trend reverses when the source comes out of the HB dip (Figure 6). The broadband noise strengthens when the HBO frequency decreases below 30 Hz (Figure 6).

The strength of HBOs and NBOs in all HID zones significantly increases with energy (Figure 5), which is consistent with previous studies (e.g., Penninx et al. 1991; Jonker et al. 2000). This trend, and particularly the fractional rms amplitude in the highest energy band of 10–20 keV, cannot be explained if the HBOs/NBOs originate solely in either the accretion disk (diskbb) or the NS surface (bbodyrad; see the right panel of Figure 2 for relative contributions of spectral components). Thus HBOs/NBOs should have a strong contribution from the Comptonizing corona (Nthcomp). However, since the disk provides seed photons to the corona, the QPO frequencies might be determined by the disk properties. Hence the oscillation could originate in the accretion disk and enhance in strength in the corona. Note that a strong dependence of the oscillation rms with the coronal properties was extensively observed in black hole X-ray binaries (e.g., Bhargava et al. 2019; Kara et al. 2019; Méndez et al. 2022; Bellavita et al. 2022; Rawat et al. 2023).

---
[9] Assuming a distance of 11 kpc (Penninx et al. 1993) and an inclination of 35° (D'Aì et al. 2009).

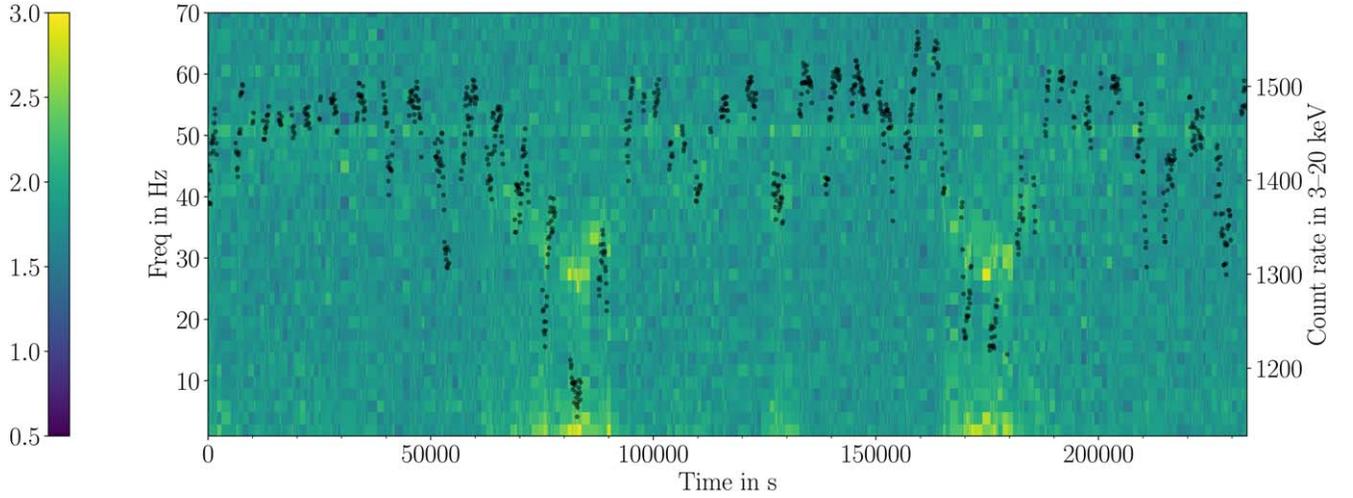

**Figure 6.** The frequency–time image of GX 340+0 with the colors indicating the variability of the source. The light curve of the source (same as in Figure 1) is shown in black points. The two deep HB excursions are clearly visible. The instrumental feature at 50 Hz is persistently present.

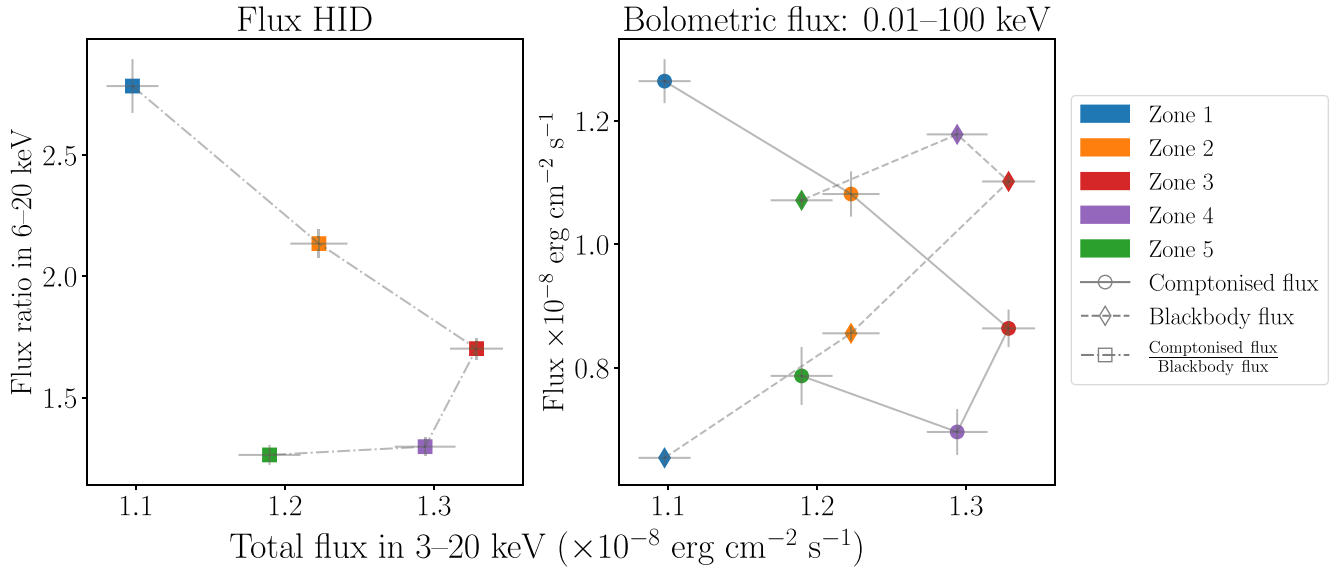

**Figure 7.** Plots involving fluxes of GX 340+0 as observed with AstroSat. Left panel: the ratio of the Comptonized flux to the blackbody flux in 6–20 keV as a function of the total flux in 3–20 keV. This panel reproduces the HB, the hard apex, and the NB of the observed Z track qualitatively (see Section 4.3). Right panel: evolution of bolometric flux (in 0.01–100 keV) values of blackbody and Comptonizing corona throughout HID zones. This panel shows that the blackbody becomes weaker and the corona becomes stronger from the hard apex to the HB (see Section 4.4).

We find that 7–20 keV photons lag 3–7 keV photons for HBOs and this lag drastically increases from zone 1 (deep HB dip) to zone 3 (hard apex; see Table 1). This could be because, as indicated by the rise of the coronal optical depth (see Figure 3), the number of Compton upscattering interactions of photons increases from zone 1 to zone 3. Such increasing scattering could also explain the decreasing strength of HBOs from the deep HB dip to the hard apex.

### 4.6. An Explanation of the Source Behavior

In our observation, GX 340+0 primarily stays around the hard apex and occasionally ventures onto HB and NB for varying distances. Since our NB excursions are short, we cannot reliably probe their origin. But, we can study the processes involved in longer HB excursions. As GX 340+0 goes onto the HB, the bolometric flux and $kT_e$ of the Comptonizing corona (Nthcomp) increase, and the bolometric flux and normalization (implying the area) of the spreading layer on the NS surface (bbodyrad) decrease (see Table 1, Figure 3, and the right panel of Figure 7). This indicates, as the source makes an excursion onto the HB, that a lesser fraction of accretion disk matter reaches the NS surface and more accreted matter replenishes and energizes the corona. This trend reverses when the source comes out of the HB. This happens for all HB excursions, regardless of their length. Thus, for longer HB excursions, the corona strengthens in ∼20 ks and weakens in a similar timescale (Figure 1).

When the corona is energized on the HB, it becomes hotter, and its physical size perhaps increases, thus covering a greater extent of the inner disk (implied by a higher `diskbb` normalization for HID zone 1; Table 1). Therefore, the corona intercepts more disk photons, and hence, its bolometric flux increases. On the other hand, the optical depth of the corona decreases (Figure 3), implying fewer Compton upscattering interactions, which makes its spectrum softer (i.e., larger `Nthcomp` $\Gamma$; see Figure 3). As mentioned in Section 4.5, these fewer upscattering interactions also explain the decreased strength and hard lag of HBOs on the HB.

## 5. Summary

Here, we summarize the main points from the study of the Z source GX 340+0 using AstroSat data.

1. As in other Cyg-like Z sources, we find that the HB and the NB appear as two different types of intensity dips from the hard apex level in 3–20 keV. Such dips can appear in any sequence and with various depths. Thus, the source moved back and forth along the track several times. In our observation, it stays close to $S_z = 1$ most of the time and makes occasional excursions onto the HB and NB.
2. The spectra that we extracted along the parts of the Z track covered by our observation can be explained by the emissions from an accretion disk, a Comptonizing corona covering the inner disk, and a blackbody emitting a spreading layer on the NS surface.
3. When the source goes into the HB, the corona is energized by the disk, and a relatively smaller amount of disk matter reaches the NS surface. Simultaneously, the coronal optical depth decreases, implying fewer Compton upscattering of a disk photon. This enhances the strength and reduces the hard lag of HBOs, which originates in the corona. This trend reverses when the source comes out of the HB dip.

## Acknowledgments

This work makes use of data from the AstroSat mission of the Indian Space Research Organisation (ISRO), archived at Indian Space Science Data Centre (ISSDC). The article has used data from the SXT and the LAXPC developed at TIFR, Mumbai, and the AstroSat POCs at TIFR are thanked for verifying and releasing the data via the ISSDC data archive and providing the necessary software tools. This research has also made use of data and/or software provided by the High Energy Astrophysics Science Archive Research Center (HEASARC), which is a service of the Astrophysics Science Division at NASA/GSFC and the High Energy Astrophysics Division of the Smithsonian Astrophysical Observatory. The authors thank Tomaso Belloni for providing the tool GHATS for timing analysis.

*Facility:* AstroSat (Singh et al. 2014).

*Software:* GHATS, XSPEC (Arnaud 1996).

## ORCID iDs

Yash Bhargava  https://orcid.org/0000-0002-5967-8399
Sudip Bhattacharyya  https://orcid.org/0000-0002-6351-5808
Jeroen Homan  https://orcid.org/0000-0001-8371-2713
Mayukh Pahari  https://orcid.org/0000-0002-5900-9785

## References

Antia, H. M., Agrawal, P. C., Dedhia, D., et al. 2021, JApA, 42, 32
Arnaud, K. A. 1996, in ASP Conf. Ser. 101, Astronomical Data Analysis Software and Systems V, ed. G. H. Jacoby & J. Barnes (San Francisco, CA: ASP), 17
Bałucińska-Church, M., Gibiec, A., Jackson, N. K., & Church, M. J. 2010, A&A, 512, A9
Bellavita, C., García, F., Méndez, M., & Karpouzas, K. 2022, MNRAS, 515, 2099
Bhargava, Y., Belloni, T., Bhattacharya, D., & Misra, R. 2019, MNRAS, 488, 720
Chakraborty, M., Bhattacharyya, S., & Mukherjee, A. 2011, MNRAS, 418, 490
Chakraborty, S., Navale, N., Ratheesh, A., & Bhattacharyya, S. 2020, MNRAS, 498, 5873
Church, M. J., Gibiec, A., Bałucińska-Church, M., & Jackson, N. K. 2012, A&A, 546, A35
Church, M. J., Halai, G. S., & Bałucińska-Church, M. 2006, A&A, 460, 233
D'Aì, A., Iaria, R., Di Salvo, T., Matt, G., & Robba, N. R. 2009, ApJL, 693, L1
Frank, J., King, A., & Raine, D. 2002, Accretion Power in Astrophysics (3rd ed.; Cambridge: Cambridge Univ. Press)
Fridriksson, J. K., Homan, J., & Remillard, R. A. 2015, ApJ, 809, 52
Gilfanov, M., Revnivtsev, M., & Molkov, S. 2003, A&A, 410, 217
Hasinger, G., & van der Klis, M. 1989, A&A, 225, 79
Hasinger, G., van der Klis, M., Ebisawa, K., Dotani, T., & Mitsuda, K. 1990, A&A, 235, 131
Homan, J., van der Klis, M., Fridriksson, J. K., et al. 2010, ApJ, 719, 201
Homan, J., van der Klis, M., Jonker, P. G., et al. 2002, ApJ, 568, 878
Homan, J., van der Klis, M., Wijnands, R., et al. 2007, ApJ, 656, 420
Iaria, R., Lavagetto, G., Di Salvo, T., et al. 2006, ChJAS, 6, 257
Jackson, N. K., Church, M. J., & Bałucińska-Church, M. 2009, A&A, 494, 1059
Jonker, P. G., van der Klis, M., Homan, J., et al. 2002, MNRAS, 333, 665
Jonker, P. G., van der Klis, M., Homan, J., et al. 2000, ApJ, 537, 374
Jonker, P. G., Wijnands, R., van der Klis, M., et al. 1998, ApJL, 499, L191
Kara, E., Steiner, J. F., Fabian, A. C., et al. 2019, Natur, 565, 198
Kuulkers, E., van der Klis, M., Oosterbroek, T., et al. 1994, A&A, 289, 795
Lavagetto, G., Iaria, R., di Salvo, T., et al. 2004, NuPhS, 132, 616
Lin, D., Remillard, R. A., & Homan, J. 2007, ApJ, 667, 1073
Lin, D., Remillard, R. A., & Homan, J. 2009, ApJ, 696, 1257
Lin, D., Remillard, R. A., Homan, J., & Barret, D. 2012, ApJ, 756, 34
Méndez, M., Karpouzas, K., García, F., et al. 2022, NatAs, 6, 577
Miller, J. M., Raymond, J., Cackett, E., Grinberg, V., & Nowak, M. 2016, ApJL, 822, L18
Misra, R., Roy, J., & Yadav, J. S. 2021, JApA, 42, 55
Mitsuda, K., Inoue, H., Nakamura, N., & Tanaka, Y. 1989, PASJ, 41, 97
Mukherjee, A., & Bhattacharyya, S. 2011, MNRAS, 411, 2717
Penninx, W., Lewin, W. H. G., Tan, J., et al. 1991, MNRAS, 249, 113
Penninx, W., Zwarthoed, G. A. A., van Paradijs, J., et al. 1993, A&A, 267, 92
Rawat, D., Méndez, M., García, F., et al. 2023, MNRAS, 520, 113
Seifina, E., Titarchuk, L., & Frontera, F. 2013, ApJ, 766, 63
Singh, K. P., Stewart, G. C., Chandra, S., et al. 2016, Proc. SPIE, 9905, 99051E
Singh, K. P., Stewart, G. C., Westergaard, N. J., et al. 2017, JApA, 38, 29
Singh, K. P., Tandon, S. N., Agrawal, P. C., et al. 2014, Proc. SPIE, 9144, 91441S
Sridhar, N., Bhattacharyya, S., Chandra, S., & Antia, H. M. 2019, MNRAS, 487, 4221
Sriram, K., Rao, A. R., & Choi, C. S. 2011, ApJL, 743, L31
van der Klis, M. 2004, arXiv:astro-ph/0410551
van Paradijs, J., Hasinger, G., Lewin, W. H. G., et al. 1988, MNRAS, 231, 379
Verner, D. A., Ferland, G. J., Korista, K. T., & Yakovlev, D. G. 1996, ApJ, 465, 487
Wilms, J., Allen, A., & McCray, R. 2000, ApJ, 542, 914
Yadav, J. S., Agrawal, P. C., Antia, H. M., et al. 2016, Proc. SPIE, 9905, 99051D
Yadav, J. S., Agrawal, P. C., Antia, H. M., et al. 2017, CSci, 113, 591
Zdziarski, A. A., Johnson, W. N., & Magdziarz, P. 1996, MNRAS, 283, 193
Życki, P. T., Done, C., & Smith, D. A. 1999, MNRAS, 309, 561